\documentclass[12pt]{article}

\usepackage[utf8]{inputenc} 
\usepackage[T1]{fontenc}    
\usepackage{hyperref}       
\usepackage{url}            
\usepackage{booktabs}       
\usepackage{amsfonts}       
\usepackage{amsmath}
\usepackage{amssymb}
\usepackage{graphicx}
\usepackage{bbm}
\usepackage{nicefrac}       
\usepackage{microtype}      
\usepackage{xcolor}         

\title{MLCBART: Multilabel Classification with Bayesian Additive Regression Trees}

\author{
  Jiahao Tian\\
  Department of Statistics and Actuarial Science\\
  Simon Fraser University\\
  Burnaby, BC, Canada V5A 1S6 \\  \texttt{jtian\_3@sfu.ca} \\[.1 in]
  Hugh Chipman \\
  Department of Mathematics and Statistics \\
  Acadia University \\ 
  Wolfville, NS, B4P 2R6 \\
  \texttt{hugh.chipman@acadiau.ca} \\[.1 in]
  Thomas Loughin \\
  Department of Statistics and Actuarial Science \\
  Simon Fraser University \\
  Burnaby, BC, Canada V5A 1S6 \\  \texttt{tloughin@sfu.ca} \\
}

\begin{document}
\maketitle

\begin{abstract}
Multilabel Classification (MLC) deals with the simultaneous classification of multiple binary labels. The task is challenging because, not only may there be arbitrarily different and complex relationships between predictor variables and each label, but associations among labels may exist even after accounting for effects of predictor variables. In this paper, we present a Bayesian additive regression tree (BART) framework to model the problem. BART is a nonparametric and flexible model structure capable of uncovering complex relationships within the data. Our adaptation, MLCBART, assumes that labels arise from thresholding an underlying numeric scale, where a multivariate normal model allows explicit estimation of the correlation structure among labels. This enables the discovery of complicated relationships in various forms and improves MLC predictive performance. Our Bayesian framework not only enables uncertainty quantification for each predicted label, but our MCMC draws produce an estimated conditional probability distribution of label combinations for any predictor values.
Simulation experiments demonstrate the effectiveness of the proposed model by comparing its performance with a set of models, including the oracle model with the correct functional form. Results show that our model predicts vectors of labels more accurately than other contenders and its performance is close to the oracle model. An example highlights how the method's ability to produce measures of uncertainty on predictions provides  nuanced understanding of classification results.


\end{abstract}

\section{Introduction}
Multilabel classification (MLC) is concerned with learning a mapping from an input space $X$ into $q$ binary labels $Y$ simultaneously.
The problem is inherently different from the multi-class classification problem in that labels are not mutually exclusive. One instance can have multiple labels with arbitrary inter-label association. Multilabel classification has received a fair amount of attention in recent years due to its increasing number of applications in various areas. For example, its use is seen in text classification, sentiment analysis, scene classification, marketing, and medical diagnosis 
\cite{ abdalrada2022machine, liu2017deep, liu2015multi,  manchanda1999shopping, maskery2008bayesian, maxwell2017deep, mehta2012multicategory}.

When multiple labels exist, there are often relationships among these labels that provide information on how they interact with one another. Leveraging this information could potentially lead to improved joint predictions. In real-world scenarios, making joint predictions is crucial across various domains. For instance, joint prediction is critical in the field of marketing. Research has revealed that certain products can serve as consumption complements to one another. By identifying these interrelationships among various products, companies can capitalize on the opportunity to bundle correlated items, thereby stimulating sales and enhancing customer value. Conversely, some products may act as consumption substitutes for others \cite{chib2002analysis}. Recognizing these substitute relationships enables marketers to develop more nuanced and effective marketing strategies. Thus, a comprehensive understanding of product relationships, whether complementary or substitutive, can significantly inform and improve overall marketing approaches, leading to optimized product offerings and increased customer satisfaction. A second application is the classification of multiple symptoms in medicine, described more fully in Section \ref{real_data}


Making joint predictions with given predictors variables is achieved through two major frameworks: problem transformation or algorithm adaptation. Problem transformation changes the goal of the analysis to something where existing techniques can be used.  For example, binary relevance \cite{tsoumakas2010mining} uses a separate binary classifier on each label while ignoring possible correlated relationships among labels. Label power-set \cite{boutell2004learning}, as the name suggests, turns the problem into a multi-class classification problem by predicting the probability for each possible label combination.  Classifier chains \cite{read2011classifier}, on the other hand, model label dependency by building a chain of binary classifiers where each classifier uses original features and predictions made by earlier classifiers in the chain. On the other hand, algorithm adaptation modifies the existing models to approach the problem by considering its unique features. These methods include Multi-Label k Nearest Neighbors \cite{zhang2007ml}, multi-class multi-label perceptron \cite{mencia2008pairwise}, and Ranking Support Vector Machine \cite{elisseeff2001kernel}.

We present a new algorithm adaptation that extends Bayesian Additive Regression Trees (BART) \cite{chipman2010bart} to explicitly model MLC data. Our proposed Bayesian model leverages the ensemble framework to produce excellent predictive performance for MLC problems and has numerous advantages over existing models and machines. Given the Bayesian nature of our model, probabilistic predictions are available. In some cases, having soft prediction and even a probabilistic distribution of labels is preferable. Our methods also model the correlation among labels naturally without changing the problem. In addition, the proposed model provides built-in methods for quantifying uncertainty in both label and probability prediction. This feature would be ideal in situations like medical diagnosis. We also present a performance analysis and comparison, as well as guidance on when and how the multivariate model is preferred over its univariate counterpart in solving the multilabel classification problem. 

The paper is structured as follows: Section \ref{background} introduces the MLC problem, notation, and the general BART model. Section \ref{methodology} presents our proposed framework and its unique features for analyzing MLC problems. In Section \ref{experiment}, we demonstrate the effectiveness of our MLCBART by comparing it against a set of other models in a small simulation experiment. Section \ref{real_data} demonstrates the application of the proposed method to real data. Section \ref{discussion} discusses how the estimation of the correlation matrix can help with MLC and other benefits offered by our proposed model. Finally, we conclude the paper in Section \ref{conclusion} by summarizing the key aspects and limitations of our work.

\section{Background \& Notation}\label{background}
Let the data be denoted by $D$ = $\{(x_i, Y_i)\;i=1,\ldots,N\}$, where $x_i$ is the vector of explantory variable of dimension $p$ of observation $i$ in the data, and $Y_{i} = (y_{i1}, \ldots, y_{iq})$ is the response vector of length $q$ where each value $y_{ik}$ is binary. 

We seek a modeling framework to predict full response vectors that achieves all of our goals---explicit modeling of correlation among labels, flexible modeling of the relationship between response and <explanatories, attributes>, and probabilistic quantification of uncertainty. 
The foundation of our framework is  the multivariate probit model, where a multivariate normal distribution (MVN) is assumed for underlying latent variables that are manifested as binary responses by thresholding \cite{chib1998analysis, edwards2003multivariate, talhouk2012efficient}. Let $Z_i = (z_{i1}, \ldots, z_{iq})$ be a vector of length $q$ denoting the underlying latent variable for instance $i$. We represent $Z_{i}$ as: 
\begin{equation}\label{formulation}
    Z_{i}=F(x_{i})+\epsilon_{i} \qquad  i=1,...,N,
\end{equation}
where $F(x_{i}) = (f_{1}(x_{i}), \ldots, f_{q}(x_{i}))$ is a vector of regression functions, and $\epsilon_i$ follows $MVN(0, \Sigma)$. The latent vector $Z_{i}$ is therefore characterized by a multivariate normal distribution with mean $F(x_{i})$ and covariance matrix $\Sigma$. Binary responses are determined according to the signs of the elements $z_{ik}$,
\begin{equation*}
    y_{ik} = \mathbbm{1} (z_{ik} > 0) \qquad k=1,\ldots, q,
\end{equation*}
where $\mathbbm{1}$ denotes the indicator function. Note that the constraints has to be put on the covariance matrix $\Sigma$ to avoid issues with parameter identifiability. Our model forces the covariance matrix to be a correlation matrix following \cite{imai2005bayesian, zhang2020parameter, zhang2006sampling}. An important characteristic of the multivariate normal model for the latent vector is that the joint prediction of $y_{ik} = 1$ can be derived by integrating the multivariate normal distribution at $f_k(x_i)$, allowing for soft classification. 

\subsection{Bayesian Additive Regression Trees (BART)}\label{BART intro}
BART is a powerful ensemble model that consists of a set of weak learners that are constrained from overfitting by a regularization prior. In the univariate case the unknown function $f(x)$ providing the mapping from predictor variables into the target variable is approximated by the sum-of-trees model $f(x) = \sum_{j = 1}^{b} g(x;T_{j},M_{j})$, where $b$ is the number of trees in the ensemble, and $g$ is a function that returns the value from the terminal node containing $x$ in regression tree $T_j$ with node values $M_j$.  Specifically, $T_j$ denotes the set of splitting rules and terminal node locations in tree $j$, and $M_j$ is a vector of parameter values with elements $\mu_{jl}$ representing the contribution to the overall mean due to terminal node $l$ in tree $j$. Each split mentioned above is based on queries of the form $x_{ia} < c$ for some contextually chosen constant $c$. Parameter values are learned probabilistically through a sequence of Markov Chain Monte Carlo (MCMC) iterations, resulting in estimated probability distribution for each parameter and any functions thereof. Each MCMC step includes modification of each of the $b$ trees by randomized choice of grow, prune or edit steps.  The sum-of-trees structure allows the BART model to capture complex nonlinear relationships. BART has been extended to various regression settings \cite{mondal2020advancement,murray2017log}. Successful adaptations of BART model in the classification setting have also been developed \cite{chipman2010bart, kindo2016multinomial, zhang2010bayesian}. Building on \cite{chipman2010bart}, which suggests binary classification using a univariate probit model, \cite{kindo2016multinomial}  handles the multiclass classification problem using a version of equation \ref{formulation}, where $F(x)$ is a vector of sum-of-trees models corresponding to different classes. We extend these ideas in a different direction to address MLC problems, and we devise a new MCMC iteration scheme that enables the estimation of posterior distributions for all mean and correlation parameters.


\section{Model and Estimation}\label{methodology}
In our model formulation, we replace $F(x_{i})$ in equation \ref{formulation} with a vector of sum-of-trees models.  Equation \ref{formulation} becomes
\begin{equation}\label{latent_mlc_bart}
    Z_{i}=G(x_{i};T,M)+\epsilon_{i} \qquad i=1,...,N,
\end{equation}
where $G(x_{i};T,M)=(G_{1}(x_{i};T,M),...,G_{q}(x_{i},T,M))$ is a vector of $q$ sum-of-tree models described in Section \ref{BART intro}.  Thus, $T$ is the parameters defining the topology of $b\times q$ trees and $M$ represents the contributions of the terminal nodes of these trees to the overall mean.  The parameters of the model, $T,M$, and $\Sigma$, are assumed to have distributions that are learned using a Bayesian algorithm as described below.
\subsection{Prior Specification}
\subsubsection{The $\Sigma$ prior}
To address the identifiability issue mentioned earlier, one possible solution is to constrain the diagonal elements of the covariance matrix $\Sigma$ to equal one. This results in $\Sigma$ in equation \ref{latent_mlc_bart} being equivalent to a correlation matrix denoted by $R$. We place a prior on an arbitrary covariance matrix $\Sigma$, and then obtain the density for the implied correlation matrix $R$ through decomposition following \cite{zhang2020parameter,zhang2006sampling}. If $\Sigma$ is a positive definite covariance matrix, it can be expressed as $\Sigma = D^{\frac{1}{2}}RD^{\frac{1}{2}}$, where $R$ is the implied correlation matrix and $D^{\frac{1}{2}}$ is a diagonal matrix of standard deviations. Then the Jacobian transformation from $\Sigma$ to the joint density of $(R, D)$ is $J_{\Sigma}(R,D) = |D|^{\frac{q - 1}{2}}$. In MLCBART, we assume that $\Sigma$ follows an inverse-Wishart distribution with degrees of freedom $m_{0}$ and scale matrix $\Sigma_{0}$ (denoted as $IW(m_0,\Sigma_0)$), with density is given by 
\begin{equation*}
    P(\Sigma) \propto |\Sigma|^{-\frac{(m_{0} + q + 1)}{2}} e^{-\frac{1}{2}tr(\Sigma_{0}\Sigma^{-1})},
\end{equation*}
where $tr$ is the trace. Then, the joint distribution of $(R,D)$ can be obtained with 
\begin{equation*}
    P(R,D) \propto |D|^{\frac{-m_{0}}{2} - 1} |R| ^{-\frac{(m_{0} + q + 1)}{2}} e^{-\frac{1}{2}tr(\Sigma_{0}(D^{\frac{1}{2}}RD^{\frac{1}{2}})^{-1})}.
\end{equation*}

\subsubsection{The $T_{kj}$ prior}
We follow the prior that is put on  the tree structure shown in \cite{chipman1998bayesian, chipman2010bart}. The probability that a given node at depth $d$ is a nonterminal node is $\frac{\alpha}{(1+d)^{\beta}}$, where $0 < \alpha < 1$ and $\beta \geq 0$ are chosen parameters. By default, the values of $\alpha$ and $\beta$ are set to 1 and 2, respectively, but these can be changed or tuned to encourage systematically larger or smaller trees. Along with this prior, there are two other priors that must be specified: (i) a uniform prior is placed on all available variables considered for a splitting rule at an internal node, and (ii) a uniform prior is assumed for all splitting values conditional on the splitting variable. 

\subsubsection{The $\mu_{kjl}\lvert T_{kj}$ prior}
The prior distribution on the contribution to the mean in terminal node $l$ for tree $j$ in label $k$ is 
\begin{equation}\label{prior_distribution}
    \mu_{kjl} \sim N(0, \sigma_{k}^{2}), \qquad k = 1,\ldots, q
\end{equation}
To choose $\sigma_k^2$, \cite{chipman2010bart} recommend assigning significant prior probability to the interval $(-3, 3)$. From equation \ref{prior_distribution}, the induced prior for $E(z_{ik} \lvert x)$ is $N(0, b\sigma_{k}^{2})$. Then $\sigma_{k}^{2}$ is chosen such that $ -r\sqrt{b}\sigma_{k}=-3$ and $r\sqrt{b}\sigma_{k}=3$, where $r$ is a tuning parameter that applies shrinkage to the means estimated by the sum-of-tree model. The value of $\sigma_k$ is then set to $3/r\sqrt{b}$. Following \cite{chipman2010bart} the default value $r=2$ is used.


\subsection{Posterior computation}
Our proposed posterior sampling scheme utilizes the Bayesian backfitting algorithm for $(T_{kj}, M_{kj})$ described in \cite{chipman2010bart}, as well as the sampling of the unbounded covariance matrix and subsequent variable transformation for the joint density $(R, D)$ \cite{zhang2020parameter}. The complete MCMC sampling scheme for each iteration is similar to the one used by \cite{kindo2016multinomial}, with a different step for sampling $R$ based on \cite{zhang2020parameter}:
\begin{enumerate}
    \item Obtain random draws from $P(Z_{i} \lvert T,M, \Sigma, Y_{i})$ for $i = 1,\ldots N$.
    \item Sample from $P((T_{kj}, M_{kj}) \lvert Z, \Sigma, Y)$ for $k = 1, \ldots, q$ and $j = 1, \ldots, b$. 
    \item Obtain a random draw from $IW(w_{p}, w_{p} \Sigma)$, where $w_{p}$ is the proposal degrees of freedom (a tuning parameter), and use the Metropolis-Hastings step to determine whether to accept it.
\end{enumerate}
  Note that at each step the most recently available parameter values are used in the computations.  Samples of the latent variable $z$ in the first step of the sampling scheme are obtained through iterative random draws of truncated univariate normals from the conditional distribution $Z_{ik} \lvert T, M, \Sigma, Z_{i(-k)} \sim N(m_{ik}, \phi_{ik})$ where $Z_{i(-k)}$ denotes the latent vector $Z_{i}$ without element $k$ and is of length $q - 1$ \cite{chib1998analysis, kindo2016multinomial}. The use of truncated normals follows from the fact that the signs of latent variables have to be compatible with observed values. The conditional mean, $m_{ik}$, and variance,  $\phi_{ik}$, of $Z_{ik}$ are given by: 
\begin{equation}
    \begin{split}
        m_{ik} &= G_{k}(x_{i},T,M)+ R_{k,-k} R_{-k,-k}^{-1}(Z_{i,-k}-G_{-k}(x_{i},T,M)), \\ 
        \phi_{ik} &= R_{k,k}-R_{k,-k}R_{-k,-k}^{-1}R_{-k,k}.
    \end{split}
\end{equation}
Here, $G_{-k}$ refers to sum-of-tree models without model $k$; $R_{k,-k}$ refers to row $k$ of $R$ without column $k$; $R_{-k, -k}$ is $R$ without row and column $k$; and $R_{-k, k}$ is the transpose of $R_{k,-k}$.
For the sampling of $(T_{kj}, M_{kj})$ for $k = 1, \ldots, q$ and $j = 1, \ldots, m$, the pseudo response $z_{ik}^{\dagger}$  for the backfitting algorithm to modify tree $j$ in the sum-of-trees model $k$ is calculated as 
\begin{equation}
        z_{ik}^{\dagger} = z_{ik} - \sum_{l \neq j}^{m}g(x_{i},T,M)- R_{k,-k} R_{-k,-k}^{-1}(Z_{i,-k}-G_{-k}(x_{i},T,M)).
\end{equation}
Each tree is viewed as a standalone Bayesian tree structure while being modified \cite{chipman1998bayesian}. Finally, the draw of the correlation matrix is obtained through the Metropolis-Hastings algorithm with the following steps: 
\begin{enumerate}
    \item generate proposal $(R^{\ast}, D^{\ast})$ by making a random draw $\Sigma^{\ast}$ from  $IW(w_{p}, w_{p} \Sigma)$;
    \item generate a new $(R, D)$ by a Bernoulli draw with acceptance probability $\min(\frac{P((R^{\ast}, D^{\ast}) \lvert G,Z, Y)\times f(\Sigma|\Sigma^{*})}{P((R, D)| G, Z, Y)\times f(\Sigma^{*}|\Sigma)},1)$.
\end{enumerate}
The proposal density $f(\Sigma^{*}|\Sigma)$ is obtained by $J_{\Sigma^{\ast}}(R^{\ast},D^{\ast}) \times IW(w_{p}, w_{p} \Sigma)$ and $\Sigma= D^{\frac{1}{2}}RD^{\frac{1}{2}}$ is the most recent MCMC draw.

\subsection{Label prediction and posterior inference}\label{inference}
The sampling scheme described above generates a sequence of $L$ MCMC draws of 
$T, M$ and $R$.  We assume these $L$ samples exclude a burn-in period.
To make predictions at some instance $x_0$, for each of the $L$ draws we randomly sample $c$ realizations of $Z$ from equation \ref{latent_mlc_bart} replacing $x_i$ with $x_0$ in $G$ and $c$ draws of $\epsilon$ from $MVN(0,R)$.
The collection of $c \times L$ samples  approximates the posterior predictive distribution for $Z$ with $c \times L$ samples for each instance. 

By thresholding each sampled latent response at zero, we can associate a particular label combination with each random sample. An estimated posterior predictive distribution is found from the proportions of samples that produce each of the $Q=2^{q}$ possible label combinations. For hard classification, we can choose the label combination with the highest posterior probability.  Alternatively, soft classification can be done using the posterior probabilities of the label combinations, allowing an implicit measure of uncertainty quantification in the label assignments.

\section{Experiment}\label{experiment} 
\subsection{Experiment setting}
The aim of the simulation study presented in this section is to study how accurately the proposed model performs multilabel classification from the perspective of MLC-specific metrics and and how well it recovers the underlying true distribution of the latent mean structure. We consider an MLC problem with $q=3$ labels and $p=3$ attributes. Each data set contains $N = 500$ training instances and $1000$ test instances. To ensure the robustness of the results, we repeat the data generation and model fitting 20 times for each scenario considered below. The attributes are sampled independently from $U(-1, 1)$. We simulate latent responses from equation \ref{formulation} with mean function $F(x)$ and correlation $R$ as given below: 
\begin{equation}\label{simu_fun}
    \begin{split}
         f_{1} &= A * \sin(\pi * X_{1} * X_{2}) - B \\ 
         f_{2} &= A * \sin(\pi * X_{1} * X_{2}) + B \\ 
         f_{3} &= A * X_{3};
    \end{split}
\end{equation}
\begin{equation}\label{cor}
R=
    \begin{bmatrix} 
    1 & 0.7& 0.8\\
    0.7 & 1 &  0.9 \\ 
    0.8 & 0.9&  1 \\ 
    \end{bmatrix} .
\end{equation}
These responses are thresholded at zero to create vectors of correlated labels.

This mean was chosen because it represents both nonlinear and interactive elements that would challenge classical multivariate linear probit models.  The strength of the signal contained in the data is controlled by parameters $A$ and $B$.  The simulation model also has fairly strong associations among labels so that we may assess how well MLCBART recovers these correlations and what effect they have on predictive performance compared to univariate methods.

Our analysis compares several models, including $q$ independent univariate BART models; a multivariate probit model  
with the correct linear predictors in each label, $\sin(\pi * X_{1} * X_{2})$ and $X_3$ provided as inputs (referred to as ``true predictors'');
a univariate probit model with true predictors; a multivariate probit model where each $f_k(x)$ is assumed linear in $X_1, X_2$, and $X_3$
(called ``linear predictors''); and a univariate probit model with linear predictors. 
The multivariate probit with true predictors is an oracle model, using information about a mean structure that we typically wouldn't have.  We expect it to perform best among all methods, both in terms of predicting $Y$ and estimating $R$.  We are interested in seeing whether MLCBART can come close to the oracle model in performance. Including the univariate version of the oracle model allows us to see what penalty, if any, is paid for ignoring correlation among labels, or if anything is gained by ignoring the correlation when it is not substantial. Similarly, the multivariate probit with linear predictors allows us to see how much performance is affected when the mean structure is incorrectly specified, and hence whether MLCBART's adaptable modeling of the latent mean offers real improvement.  Finally, univariate (binary relevance) versions of BART and probit with linear predictors provide a more complete assessment of the relative importance of mean modeling versus variance modeling.

\subsection{Experiment results analysis}
In our analysis, we consider three different MLC-specific metrics---subset accuracy, macro F1 score, and macro precision \cite{madjarov2012extensive}---to demonstrate the performance of the models considered in the experiment. Subset accuracy refers to the proportion of predicted sets of labels that match the observed set of labels. This is a rather strict measure as a prediction for an observation is considered correct only if all labels are predicted correctly for that observation. We expect that the model's performance on this measure could be improved by using the correlation matrix to capture the co-occurrence pattern that is unexplained by predictor variables. The two other metrics are aggregate measures that summarize the label-wise F1 score and precision across all labels, providing a comprehensive summary of the models' performance.

With multiple replications in each simulation scenario, we can perform formal analysis of variance (ANOVA) comparisons of the models' performance. 
The ANOVA model that we fit includes a block effect to capture variability across data sets that is common to all methods and the model effect to capture the differences in performance results from the models.   
The results indicate that the model effect is statistically significant for each metric, leading to the conclusion that there are differences in performance among the models. 

\begin{figure}[ht]
    \centering
    \includegraphics[width=\textwidth]{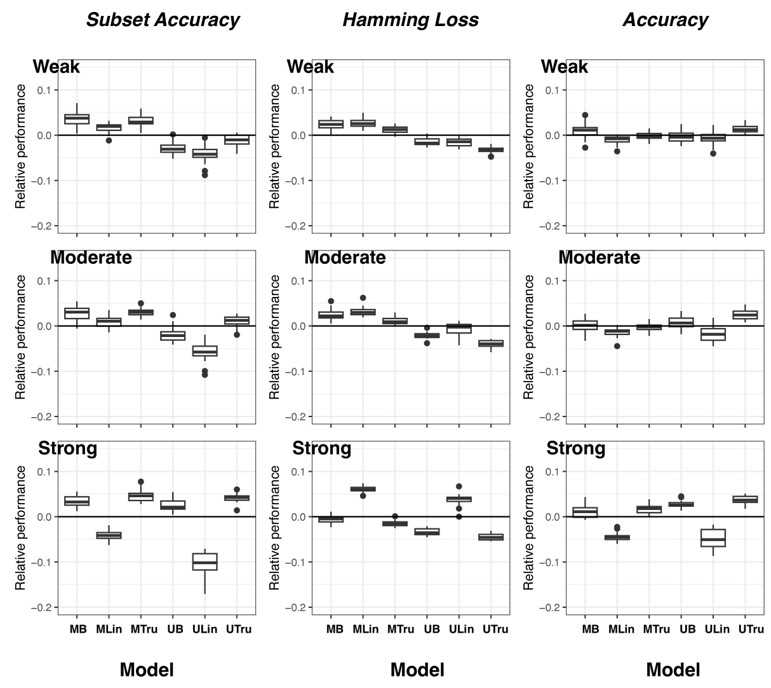}
    \caption{The comparative performance of different models under subset accuracy, macro F1, and macro precision metrics on data sets with weak, moderate, and strong signals. Boxes summarize relative (centered) performance of each model across 20 data sets. The models included in the evaluation were Multivariate BART (MB), Multivariate probit with linear predictors (MLin), Multivariate probit with true predictors (MTru), Univariate BART (UB), Univariate probit with linear predictors (ULin), and Univariate probit with true predictors (UTru).}
    \label{simu_res}
\end{figure}

To explore the differences more deeply, we center the models' subset accuracy results around zero for each data set, which allows us to remove the effects of different data sets and compare relative performance of the models directly.  Figure \ref{simu_res} shows these centered model effects from each metric (labeled as ``relative performance'') for each model under values of $A$ and $B$ that lead to weak, moderate, and strong signals.

The first column of Figure \ref{simu_res} represents the evaluation of the model's performance on subset accuracy. Subset accuracy is a measure that only considers predictions to be correct when the predicted labels for all margins are correct. It is expected that obtaining an accurate correlation estimate is crucial to have a satisfactory performance on this measure.   In the first row of the plot, the coefficients $A$ and $B$ are set to 0.3 and 0.1, respectively, resulting in a probability of the label being 1 ranging from approximately 0.3 to 0.6 for all margins. This configuration places the latent variables close to the decision boundary. In this setting, all of the multivariate models perform better than their univariate counterparts. This suggests that accurately estimating the conditional correlation matrix is crucial in this scenario. Our proposed model achieved a similar performance to the best possible model which is not available in the real world.



The last row of the first column of the plot corresponds to the case where the data contains relatively strong signals, with coefficients $A$ and $B$ set to 1 and 0.1, respectively, causing the probability of a label being 1 to range from 0.1 to 0.9. By comparing the performance of multivariate model with their univariate counterparts, there is no significant difference in terms of model performance. This suggest that mean structure has more impact on the final performance than the estimation of the correlation matrix. This does not come as a surprise since we are estimating the correlation matrix of the residual error. When the latent mean is far from the decision boundary, the residual error cannot flip the label. In addition to that, it's also more difficult for the model to produce an accurate estimate of the correlation structure because there is not enough information in the data. Even in this case, our proposed model still achieved performance close to the best model. Also note that, our proposed model performed similar to the univariate model even if our model has unwanted complexity in this case.

\begin{table}[ht]
\centering
\begin{tabular}{|c|c|c|}
\hline
\textbf{True Correlation Coefficient} & \textbf{Weak Signal} & \textbf{Strong Signal} \\
\hline
0.7 & 0.63 $\pm$ 0.026 & 0.35 $\pm$ 0.088 \\
\hline
0.8 & 0.76 $\pm$ 0.038 & 0.46 $\pm$ 0.092\\
\hline
0.9 & 0.82 $\pm$ 0.035 & 0.52 $\pm$ 0.078 \\
\hline
\end{tabular}
\caption{Mean and Standard deviation of the posterior distribution of  Correlation Coefficient in the case of weak and strong signal across twenty repetitions}
\label{cor_table}
\end{table}

The correlation matrix shown in equation \ref{cor_table} contains the posterior mean and the corresponding standard error over twenty iterations of simulation. It turns out that our proposed model is able to get a reasonably accurate estimate of the underlying correlation matrix. What is also worth mentioning is that the standard error is quite small, suggesting that the model is confident about the estimate. However, in the case of strong signals, we observed that the posterior mean deviated significantly from the true underlying correlation coefficient. Additionally, the standard error increased compared to the weak signal scenario, suggesting that the model encountered difficulties in identifying a high posterior probability region for the correlation matrix. This phenomenon can be attributed to the fact that the data in the strong signal case contains less informative variation, making it more challenging for the model to estimate the correlation accurately. What's worth pointing out that this is not specific to MLCBART. Instead, it's an identificability problem that are evident in all other multivariate methods considered. 

The above analysis examines performance in terms of subset accuracy, a multivariate measure \cite{madjarov2012extensive}. 

One unique feature of our proposed model is the ability to obtain the posterior predictive distribution of the label combinations. As described in the Section \ref{inference}, we use a repeated sampling approach to obtain counts for each label combination, which are then converted into probabilities. 
The estimated probability distribution for the univariate BART model is obtained by assuming that the joint probability of the label combination is the product of the probabilities of each individual label. To evaluate the accuracy of these estimated distributions, we use the Kullback-Leibler (KL) divergence to measure the distance between the estimated distribution and the true underlying distribution from the respective simulation models.  We computed KL divergence comparing distributions of label combinations for each instance in each test set. We averaged these measures within each test set, and computed summary statistics on these averages.  The results are in Table \ref{KL_table}.  Note that smaller values indicate a greater degree of similarity between the estimated and true distributions.

\begin{table}[ht]
    \centering
    \begin{tabular}{ |c|c|c|c| }
     Model & Mean & Median & Setting\\ 
     \hline
     Multivariate BART & 0.0584 & 0.0555 & Weak signal \\ 
     Univariate BART & 0.4852 & 0.4840  & Weak signal \\
     Multivariate BART & 0.1010 & 0.1000 & Strong signal \\ 
     Univariate BART & 0.4075 & 0.4050 & Strong signal \\ 
    \end{tabular}
    \caption{Summary statistics of Kullback-Leibler divergence for the multivariate BART and univariate model for all simulation settings}
    \label{KL_table}
 \end{table}

The table shows that the estimated probability distribution given by the multivariate model is substantially closer to the underlying distribution compared to the univariate BART model. This finding indicates that the proposed model is more capable of uncovering underlying relationships from the distribution perspective than its univariate counterpart by accurately estimating both the mean structure and the correlation matrix.

In addition to the observed improvements in the aforementioned metrics, it is of interest to examine whether MLCBART is capable of accurately uncovering the underlying conditional correlation matrix $R$ given in equation \ref{cor}. We extracted this information for one simulated data set.  The true correlations 0.8, 0.7, and 0.9 had posterior distributions of correlation values with means 0.821, 0.762, and 0.877, with first and third quartiles (0.805, 0.841), (0.728, 0.799), and (0.847, 0.900), respectively. 

We were interested in how our proposed model would perform when presented with data containing additional predictors that have no predictive power. To investigate this, we added two extra predictors that follow a $U(-1, 1)$ distribution but do not influence the means in any of the three simulation settings discussed earlier. We analyzed the performance of all models under these conditions. The results indicate that the presence of additional variables in the training data had no observable impact on the performance of our proposed model. This suggests that our model, with its flexible tree structure, is capable of identifying variables with predictive power and disregarding those that are not useful for the predictive task.

This experiment primarily showcases the effectiveness of MLCBART in terms of metrics that are specific to MLC, such as the estimated correlation matrix, the estimated probability distribution (measured by KL divergence). It is encouraging that MLCBART also performs well on metrics that ignore correlation and focus only on classifying individual labels. 
\section{Real Data Analysis}\label{real_data}
In this section, we present an analysis of empirical data derived from \cite{shao2013symptom}. The dataset comprises records pertaining to coronary heart disease, encompassing a total of 555 cases. The sample population consists of 265 male patients (47.7\%) and 290 female patients (52.3\%). The study examines five distinct syndromes and incorporates 49 predictor variables.
It is important to note that the original dataset contained six labels. However, one column exhibited a perfect separation issue, necessitating its exclusion from our analysis. Consequently, we focus our subsequent examination on the remaining five labels.

To evaluate the efficacy of the proposed methodologies, we employed a rigorous cross-validation approach. The dataset was partitioned into an 80\% training set and a 20\% test set. The model was then fitted to the training data and subsequently evaluated on the test set. To mitigate potential bias arising from random data partitioning, this process was iterated ten times. For each iteration, we conducted a comparative analysis between our proposed method and a univariate Bayesian Additive Regression Trees (BART) model. The selection of the univariate BART model as a benchmark was predicated on its robust performance in the simulation study detailed in the preceding section. Additionally, we incorporated a naive baseline model that consistently predicts the most frequently occurring label combination observed in the training set. The aggregated performance metrics for each method under consideration are summarized in Table \ref{acc_table}. This comprehensive evaluation framework allows for a more robust assessment of the proposed method's effectiveness relative to established alternatives.

\begin{table}[ht]
    \centering
    \begin{tabular}{ |c|c|c| }
     Model & Subset Accuracy & Accuracy \\ 
     \hline
     Multivariate BART & 0.19 & 0.56 \\ 
     Univariate BART & 0.18 & 0.56   \\
     Baseline & 0.14 & 0.53 \\ 
    \end{tabular}
    \caption{Performance of various models on repeated data split of CHD data}
    \label{acc_table}
 \end{table}

Our analysis revealed a statistically significant difference in subset accuracy between the MLCBART and Univariate BART models. This disparity suggests the presence of unexplained variance within the dataset that cannot be adequately accounted for by the predictor variables utilized in the univariate model.
This interpretation is corroborated by an examination of the posterior mean of the estimated correlation coefficients. We observed moderate to strong correlations among certain labels, indicating complex interdependencies within the label space. Specifically:
\begin{itemize}
    \item Labels one and two exhibited a strong negative correlation, with a coefficient of -0.85.
    \item Labels two and three demonstrated a moderate positive correlation, with a coefficient of 0.54.
\end{itemize}
These findings elucidate the superior performance of the multivariate model in terms of subset accuracy on this particular dataset. The ability of the MLCBART model to capture and leverage these inter-label correlations likely contributes to its enhanced predictive capabilities compared to the univariate approach. This observation underscores the importance of considering label dependencies in multi-label classification tasks, particularly when dealing with complex, real-world datasets where such relationships are prevalent. Future research may benefit from further exploration of these label interactions and their impact on model performance across various domains.

\subsection{Uncertainty Quantification}
\begin{figure}[ht]
    \centering
    \includegraphics[width=\textwidth]{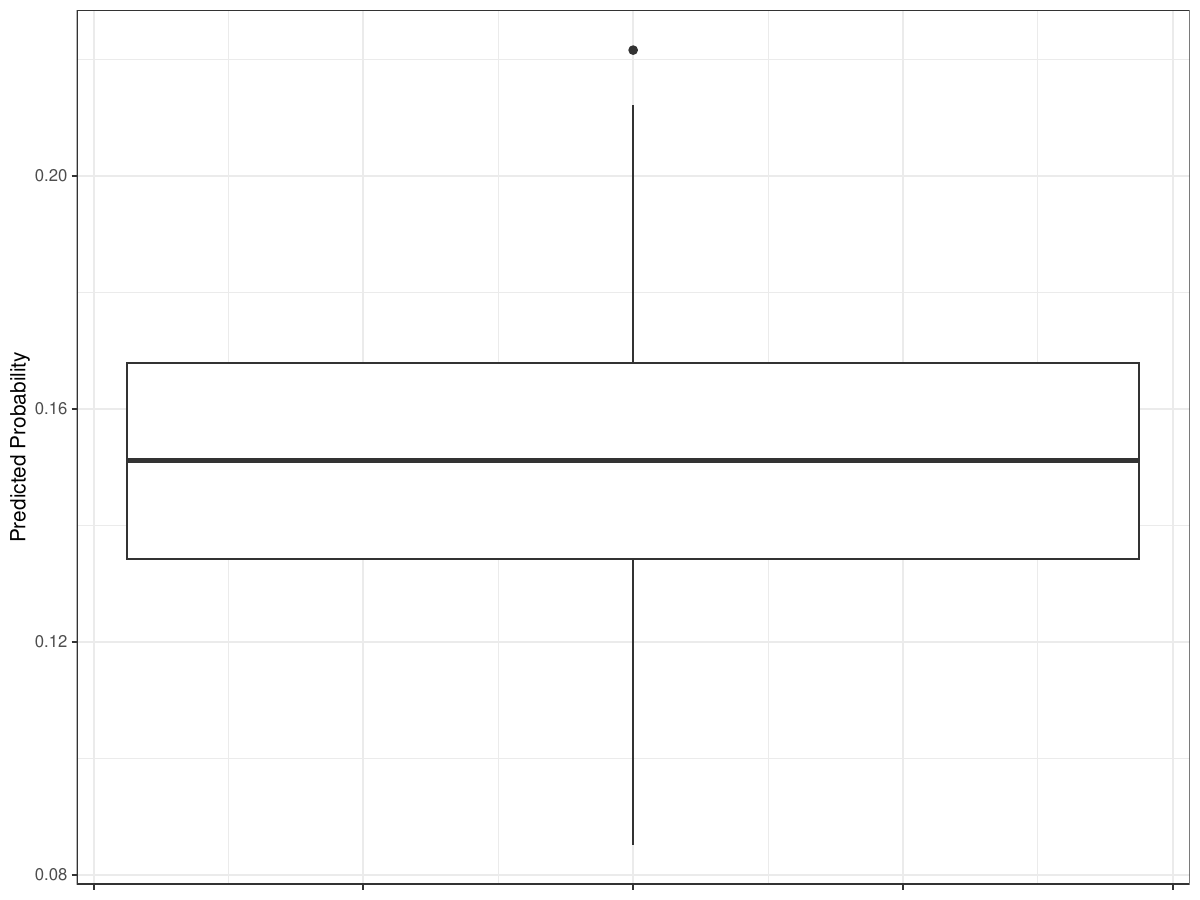}
    \caption{Boxplot of top probability for a certain data split}
    \label{top_prob}
\end{figure}

A distinctive advantage of our proposed method MLCBART, rooted in the Bayesian framework, is its capacity to provide comprehensive uncertainty quantification for each predicted label combination. Unlike traditional approaches, our method assigns a probability to each potential label combination, offering a nuanced perspective on the model's predictions. For example, Figure \ref{top_prob} illustrates the distribution of the highest probabilities assigned to label combinations for each example in the dataset. The box plot reveals that the model's confidence in its predictions is relatively low, with the highest probabilities clustering around 20\%. This observation suggests two important insights:
\begin{itemize}
    \item The model exhibits a degree of uncertainty in its predictions, which is valuable information for decision-making processes.
    \item The dataset under analysis presents significant challenges for accurate modeling, indicating its complexity and the potential need for more sophisticated approaches or additional informative features.
\end{itemize}

As previously discussed, it is feasible to derive an empirical distribution of label combinations from the posterior distribution for each example. Table \ref{pred_real_data} illustrates this distribution for a specific instance. The table demonstrates a key advantage of our proposed framework for uncertainty quantification: the ability to quantify uncertainty both across label combinations (row-wise) and for individual labels (column-wise). In the presented example, the model identifies $(1,0,1,1,1)$ as the most probable label combination, albeit with a degree of uncertainty. At the individual label level, the model exhibits high confidence in predicting a value of one for labels one, three, and five. However, it demonstrates considerable uncertainty regarding the predicted value for label three.


\begin{table}[ht]
\centering
\begin{tabular}{|c|c|c|c|c|c|c|}
\hline
Rank & L1 & L2 & L3 & L4 & L5 & $\hat{P}(Y=y)$ \\
\hline
1 & 1 & 0 & 1 & 1 & 1 & 0.222 \\
\hline
2 & 1 & 0 & 1 & 0 & 1 & 0.119 \\
\hline
3 & 1 & 0 & 0 & 1 & 1 & 0.113 \\
\hline
4 & 1 & 0 & 0 & 0 & 1 & 0.078 \\
\hline
5 & 0 & 0 & 0 & 1 & 1 & 0.066 \\
\hline
{\vdots} & {\vdots} & {\vdots} & {\vdots} & {\vdots} & {\vdots} & {\vdots} \\
\hline
 $\hat{P}(Y_{q}=1)$& 0.669 & 0.152 & 0.521 & 0.602 & 0.823 &  \\
\hline
\end{tabular}
\caption{Probabilities for different label combination for a particular example}
\label{pred_real_data}
\end{table}

\section{Discussion}\label{discussion}

We have proposed a multivariate semi-parametric model for MLC that assumes a correlated multivariate normal distribution for a latent vector of variables, which is then transformed into a vector of binary values through a threshold function. In addition to estimating the means in the latent space through $G(X;T,M)$, our model also estimates the conditional correlation matrix through posterior draws and uses it to predict {\em combinations} of labels. On the other hand, the univariate BART model assumes that each label is independent from each other after taking into account predictor variables, which is equivalent to an MLCBART model with an imposed identity correlation matrix.   We discuss some of the advantages provided by the multivariate structure and its flexible correlation matrix.  
\subsection{Conditional correlation matrix}\label{conditional_cor}
The MLCBART model does not impose any structure or other assumptions on $R$, aside from its existence as a positive definite matrix.  This is very useful because, conditional on the covariates, pairs of labels can be either positively associated or negatively associated, and both types may occur arbitrarily within a given set of labels. For example, certain diseases may tend to occur in the same people (co-morbidity), resulting in positively correlated labels for these diseases.  On the other hand, labels for weather phenomena may exhibit both positive and negative correlations, such as ``cloudy'', ``snowy'', and ``hot''.  These correlations could be innate to the label constructs, or they could be partially or entirely driven by unknown or unobserved features that relate to the labels in similar or different ways.  By estimating $R$ we are essentially estimating the aggregate effects of all unobserved factors that impact the co-occurrence of labels.  As long as the model is applied to future data that are drawn from the same population and have the same relationships between labels and unmeasured features, the estimation of $R$ provides an opportunity to incorporate more information into the prediction process than what is available from the measured predictors.  

In a similar way, model misspecification results in errors of bias that also fall into the correlation estimates after models for the means are fitted.  MLCBART uses an established tree-based ensemble model for the latent mean, and hence can be expected to be quite flexible and adapt any true mean structure.  Nonetheless, systematic imperfections in the mean model could occur.  If they do, it is possible that correlation matrix would also provide some cushion against poor performance, as we observed in the two models with linear predictors and a strong mean signal, although we have not yet investigated this in the context of MLCBART. 

We have observed an interesting relationship between strength of mean signal and value of modeling the conditional correlation matrix, which is partly shown in Figure~\ref{simu_res}. When the signal in the data is not strong, the effects captured by $R$ from unobserved/unknown factors become essential in determining the label combinations. Therefore, incorporating these effects in the predictions using our novel resampling scheme yields significant performance gains for the MLCBART. However, in cases where the mean signal in the data is strong, the importance of correlation is overshadowed by the importance of estimating the mean structure as closely as possible. Again, MLCBART's sum-of-trees model should allow it to adapt to a very broad range of mean structures, enabling strong performance in many real problems.

\subsection{Uncertainty quantification}
MLCBART can do soft classification using probabilities.  This is an inherent quantification of uncertainty for hard classification, and allows an analyst to distinguish instances where classification is relatively certain from those where it is not.  But we can go further because of the Bayesian framework.  MLCBART can generate probability estimates for each label with each MCMC draw.  So if soft classification is of more explicit interest, we can find credible intervals for each label's probability and use these quantify the uncertainty in the marginal probability estimates.

For predicting entire label sets---or even partial label sets---we can quantify uncertainty through the posterior predictive distribution of label sets as described in Section \ref{inference}.  Through the MCMC draws we learn not only which combination of labels is most probable, but also its probability and the probabilities of any other label combination.  This may suggest the potential for using alternative label sets with estimated probabilities close to the maximum when there is some compelling reason to do so.   

Selecting the labels with the highest posterior probability implicitly assumes that misclassification (0-1) loss is being used.  However, other loss functions may be preferred, including functions with unequal weights on different forms of misclassification.  Because MLCBART produces an estimate of the entire distributions of labels and label combinations, it is easy to apply different loss functions \it a posteriori.\rm  The posterior expected loss associated with any label choice is found by simply calculating its loss against each observed label or combination of labels in the posterior, and then using their estimated probabilities to compute the posterior expectation.   

\subsection{Joint prediction versus individual prediction}
In the preceding section, we demonstrated that our proposed method achieves superior performance in joint predictions compared to all other methods under consideration. While measures such as subset accuracy favor approaches that leverage inter-label relationships, we also evaluated other metrics, including hamming loss, to provide a comprehensive assessment. Interestingly, our proposed model does not consistently outperform others across all metrics. For instance, in terms of hamming loss, univariate methods generally exhibited better performance, indicating their ability to produce more partially correct predictions. This observation is not unique to our proposed method but applies to all multivariate approaches examined in this study.
Consequently, if the primary objective is to maximize the number of correct predictions for each individual label, univariate methods would be the preferred choice. However, multivariate methods, including our proposed approach, are more suitable when the goal is to make accurate joint predictions that account for label dependencies.

\section{Conclusion}\label{conclusion} 

We propose a Bayesian framework for handling the MLC problem and demonstrate its many appealing features through simulation studies. The essential components of our model are the individual sum-of-trees model and the estimation of the conditional correlation matrix. The non-parametric and flexible Bayesian trees allow for the discovery of complex relationships among the data, while the conditional correlation matrix boosts the predictive performance for MLC. Our Bayesian model makes it possible to quantify uncertainties for both label levels and label combinations. The estimated probabilistic distribution for each instance allows the prediction to be set based on costs. The estimation of the conditional correlation matrix not only enhances the model's predictive performance but also provides a qualitative assessment of correlation among labels. The experiment indicates that our proposed model is able to uncover the underlying $R$ quite accurately.  

There are some limitations to this approach. MLC BART is relatively computationally expensive and is most suitable for small- to moderate-sized problems. We have not yet attempted to develop a parallelized version that might be able to analyze larger problems. We also have not yet explored the sensitivity of the model to different priors and tuning parameters.  However, we are encouraged by its strong performance using default settings in our limited simulation and recognize that effective tuning could make it even better, albeit at a high computational cost. We have investigated the method's performance on just a small simulation under one basic structure.  While this structure was chosen to challenge MLCBART is several ways, a more complete analysis of model performance would be welcome.  Finally, we have not compared MLCBART to any of the other recently-developed methods specifically for MLC prediction.  It would certainly be useful to see whether our model can outperform existing methods over a broad range of problems.

Despite these limitations, our model's ability to produce performance close to the oracle model in the simulation study and to produce uncertainty quantification makes it a strong contender for solving MLC problems.


\bibliographystyle{elsarticle-num} 
\bibliography{bibiliography}
\end{document}